\documentclass[12pt,preprint]{aastex}

\begin{document}

\title {Detection of a planetary system orbiting the eclipsing polar HU
Aqr}

\author {Qian S.-B.\altaffilmark{1,2,3}, Liu
L.\altaffilmark{1,2,3}, Liao W.-P.\altaffilmark{1,2,3}, Li
L.-J.\altaffilmark{1,2,3}, Zhu L.-Y.\altaffilmark{1,2,3}, Dai
Z.-B.\altaffilmark{1,2}, He J.-J.\altaffilmark{1,2}, Zhao
E.-G.\altaffilmark{1,2},, Zhang J.\altaffilmark{1,2} and Li
K.\altaffilmark{1,2,3}}

\altaffiltext{1}{National Astronomical Observatories/Yunnan
Observatory, Chinese
     Academy of Sciences, P.O. Box 110,
     650011 Kunming, P. R. China (qsb@ynao.ac.cn)}

\altaffiltext{2}{Key laboratory of the structure and evolution of
celestial
  objects, Chinese Academy of Sciences, P.O. Box 110, 650011
  Kunming, P. R. China}

\altaffiltext{3}{Graduate University of the Chinese Academy of
Sciences, 100049 Beijing, P. R. China
 Yuquan Road 19\#}

\begin{abstract}
Using the precise times of mid-egress of the eclipsing polar HU Aqr,
we discovered that this polar is orbited by two or more giant
planets. The two planets detected so far have masses of at least 5.9
and 4.5\,${M_{Jup}}$. Their respective distances from the polar are
3.6\,AU and 5.4\,AU with periods of 6.54 and 11.96 years,
respectively. { The observed rate of period decrease derived from}
the downward parabolic change in O-C curve is { a factor 15 larger
than the value expected for gravitational radiation}. This indicates
that it may be only a part of a long-period cyclic variation,
revealing the presence of one more planet. It is interesting to note
that the two detected circumbinary planets follow the Titus-Bode law
of solar planets with n=5 and 6. { We estimate that another 10 years
of observations will reveal the presence of the predicted third
planet.}
\end{abstract}

\begin{keywords}
          Stars: binaries : close --
          Stars: binaries : eclipsing --
          Stars: individuals (HU Aqr) --
          Stars: white dwarf --
          Stars: planetary system
\end{keywords}

\section{Introduction}

With an orbital period of 2.08 hours, HU Aqr is a member of the AM
Her subclass (also called polars because of their highly polarized
optical emission) of cataclysmic variables (CVs) with the magnetic
field of the white dwarf primary strong enough to prevent materials
from the main-sequence companion for forming an accretion disc
(Warner 1995). It is the brightest among 15 known eclipsing polars
at both optical and X-ray wavelengths (Schwope et al. 1993, 2001).
Since the discovery by ROSAT in 1993, it has become one of the most
comprehensively observed polars in various wavelength bands (Schwope
et al. 1993, 2001; Harrop-Allin et al. 1999; Bridge et al. 2002;
Schwarz et al. 2009).

Using the 2.4-m optical telescope in Lijiang station of Yunnan
Astronomical Observatory, we monitored HU Aqr for nearly one year
(from May 20, 2009 to May 17, 2010), and 10 eclipse profiles were
obtained. We discovered that the ingress shapes of those eclipses
are variable, while the profiles of the eclipse egress were as
stable as those observed at other wavelengths (e.g., X-ray). It is
therefore possible to register very small differences in the arrival
times { of the mid-egress photons at various wavelengths}, allowing
the detection of extremely low-mass circumbinary objects through the
analysis of the observed-calculated O-C diagram ("O" refers to the
Observed times of the eclipse egress, while "C" to those Computed
with a linear ephemeris). This method is similar to the radio
approach used to detect planets around pulsars (e.g., Backer et al.
1993), { and has been used to find giant planets orbiting the other
eclipsing polar DP Leo (Qian et al. 2010; Beuermann et al. 2011a),
the extreme horizontal branch pulsating star V391 Peg (Silvotti et
al. 2007), and the others listed by Silvotti et al. (2010).}

\section{New mid-egress times and the changes of the O-C diagram}

The orbital period of the brightest eclipsing polar HU Aqr was first
noticed to be variable by Schwope et al. (2001). Recently, 72 times
of accretion spot egress in optical, UV and X-rays were measured and
collected (Schwarz et al. 2009). It is shown that the O-C diagram of
the observed accretion spot eclipse timings reveals complex
deviations from a linear trend, and a constant or cyclic period
change or a combination thereof cannot describe the general O-C
trend. { To understand the properties of the O-C variation, HU Aqr
was monitored from} May 20, 2009 using a VersArray 1300B CCD camera
mounted on the 2.4-m telescope at Lijiang station of Yunnan
Astronomical Observatory. During the observation, no filters were
used. In all, 10 complete eclipses were observed. Examples of these
eclipses are displayed in Fig. 1.

\begin{figure}
\begin{center}
\includegraphics[angle=0,scale=1.1]{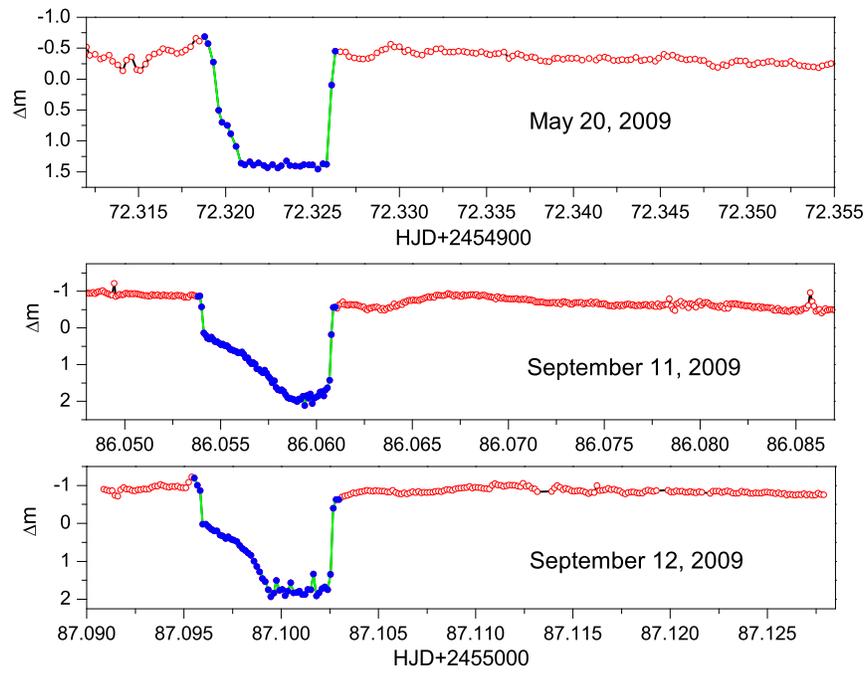} \caption{{
No-filter light curves} of HU Aqr obtained with the 2.4-m telescope
from May 20 to September, 2009. Blue solid dots refer to the eclipse
profiles, while red circles denote out of eclipses observations.}
\end{center}
\end{figure}

{ The no-filter light curves} displayed in Fig. 1 show a number of
features that are characteristic of an eclipsing polar system. The
eclipse starts with the limb of the secondary star (the red dwarf
component) eclipsing the accretion region, and the white dwarf is
also eclipsed nearly at the same time. Then the accretion stream is
the dominant source of the brightness with a small contribution of
the red dwarf component, and finally only the secondary star is
visible and provides a constant contribution. The sequence of the
egress is approximately reversed. As shown in Fig. 1, the profiles
of the ingress vary from one to another, while those of the egress
are very stable. When we started to monitor the binary on May 20,
2009, it was entering into a high-accretion state and was in a high
state on Sep. 11 and 12, 2009. Since the profiles of the egress are
very stable, the times of mid-egress can reliably be used for period
investigation. Moreover, HU Aqr is one of the brightest eclipsing
polars at optical and X-ray wavelengths. These properties make HU
Aqr the most suitable binary system for determining high-precision
period changes.

{ We defined as egress times those with half intensity (mid-egress),
as it was done by Schwarz et al. (2009).} The integration time for
each CCD image was 5s, and the readout time of the VersArray 1300B
CCD camera was about 1.8s. Therefore, the time resolution of the
photometric data is about 6.8s. We estimate that the error of the
timings of mid-egress is about half of the time resolution, i.e.,
3.4\,s=0.000039\,days. Since UTC (coordinated universal time) is
affected by the fluctuations of the Earth's rotation and is not
uniform, the determined eclipse egress times have been converted to
Barycentric Dynamical Time (TDB). The BJDs listed in Table 1 are
barycentrilly-corrected TDB, while HJDs are the heliocentric UTC
times. Therefore, their differences are close to 60\,s (e.g.,
Eastman et al. 2010).

\begin{table}
\begin{center}
\caption{New eclipse egress times of the eclipsing polar HU Aqr.}
\begin{tabular}{llll}\hline\hline
E &  HJD (days) & BJD(days) &  Errors (days)\\\hline
67604 & 2454972.326050 & 2454972.326789 & 0.000039\\
68914 & 2455086.060739 & 2455086.061477 & 0.000039\\
68926 &  2455087.102611& 2455087.103349 & 0.000039\\
69328 & 2455122.004406 & 2455122.005145 & 0.000039\\
69490 &  2455136.069318& 2455136.070057 & 0.000039\\
69800 & 2455162.983604 & 2455162.984342 & 0.000039\\
69812 & 2455164.025454 & 2455164.026193 & 0.000039\\
69823 & 2455164.980456 & 2455164.981195 & 0.000039\\
69915 & 2455172.967933 & 2455172.968672 & 0.000039\\
71785 & 2455335.322021 & 2455335.322762 & 0.000039\\
\hline
\end{tabular}
\end{center}
\end{table}

Those mid-egress times collected by Schwarz et al. (2009) are in
BJED, the barycentrilly corrected ephemeris time (ET). Since 1
January, 1984, ET was replaced by TDT (Terrestrial Dynamical Times)
and the difference between TDT and TDB is no more than 0.0017\,s.
Therefore, BJEDs are actually the same as the BJDs. In order to
compare with the O-C diagram published by Schwarz et al. (2009), the
same linear ephemeris,
\begin{equation}
BJD = 2449102.9201788+0.08682040612\times{E},
\end{equation}
used by them was applied to calculate the O-C values. To construct
the diagrams, 72 eclipse egress times from the literature (blue) and
10 new ones (red) were used. The corresponding O-C diagram is
displayed in the top panel of Fig. 2. The investigation by Schwarz
et al. (2009) revealed that a long-term decrease or a cyclic change
or a combination of them cannot describe the general O-C trend well.
Therefore, it seems that there are two cyclic variations in the O-C
curve that are caused by a pair of light-travel time effects via the
presence of two companions { HU Aqr (AB)b and HU Aqr (AB)c)}. To
describe the O-C curve well, a combination of a quadratic ephemeris
and two additional periodic terms are required,
\begin{equation}
O-C=\Delta{T_{0}}+\Delta{P_{0}}\times{E}+\frac{\beta}{2}E^{2} +
{\tau}_A + {\tau}_B,
\end{equation}
where $\Delta{T_{0}}$ and $\Delta{P_{0}}$ are the revised epoch and
period respect to the ephemeris values in Eq. (1), $\beta$ is the
rate of the linear period decrease, and ${\tau}_A$ and ${\tau}_B$
are the two cyclic changes.

Our best fit to the O-C diagram (the solid magenta line in the top
panel of Fig. 2) reveals that the orbit of HU Aqr (AB)b is circular,
while that of HU Aqr (AB)c is eccentric (e.g., Irwin 1952), i.e.,
\begin{equation}
{\tau}_A=K_A\sin(2\pi/P_A\times{E}+\varphi),
\end{equation}
and
\begin{eqnarray}
{\tau}_B&=&K_B[(1-e^{2})\frac{\sin(\nu+\omega)}{1+e\cos\nu}+e\sin\omega]\nonumber\\
&=&K_B[\sqrt{1-e^{2}}\sin{E^{*}}\cos\omega+\cos{E^{*}}\sin\omega],
\end{eqnarray}
where $\nu$ is the true anomaly, $E^{*}$ is the eccentric anomaly,
$K_A=\frac{a_A\sin{i_A}}{c}$, and $K_B=\frac{a_B\sin{i_B}}{c}$
($a_A\sin{i_A}$ and $a_B\sin{i_B}$ are the projected semi-major axes
and c is the speed of the light). In solving ${\tau}_B$, the two
correlations,
\begin{equation}
N=E^{*}-e\sin{E^{*}},
\end{equation}
and {
\begin{equation}
N=2\pi(t-T)/P_{B}
\end{equation}
}were used, where $N$ is the mean anomaly and $t$ is the time of
mid-egress. The other parameters and the derived values are
described in Table 2. The light travel-time effect amplitudes of HU
Aqr (AB)b and HU Aqr (AB)c are 9.2\,s and 10.5\,s, respectively. The
derived orbital periods are 6.54 years for HU Aqr (AB)b and 11.96
years for HU Aqr (AB)c.

\begin{figure}
\begin{center}
\includegraphics[angle=0,scale=1.1]{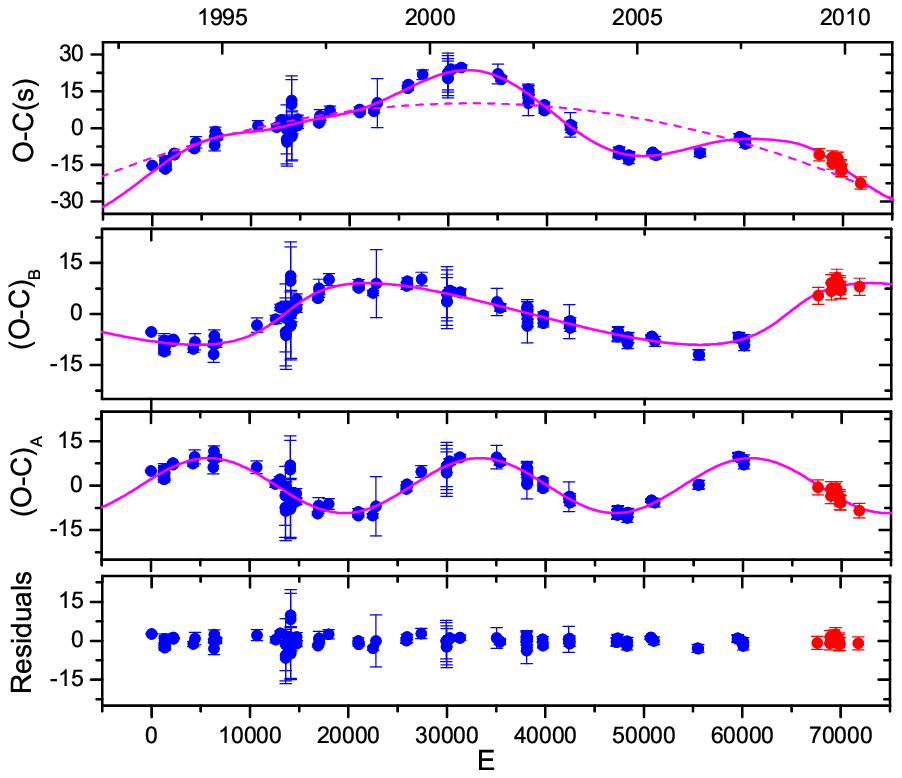} \caption{O-C
diagrams of the eclipsing polar HU Aqr. Blue dots refer to the data
compiled from the literature, while red ones to our new
observations. It is shown in the top panel that a combination (solid
magenta line) of two cyclic variations and a linear decrease (dashed
magenta line) can give a good fit to the general trend of the O-C
data. The two cyclic changes (Case A: $P_A=6.54$ years and
$K_A=9.25$\,s; Case B: $P_B=11.96$ years and $K_B=10.54$\,s) are
displayed in the middle panels. After all of the variations were
removed, the residuals are displayed in the lowest panel where no
variations can be traced there.}
\end{center}
\end{figure}

\begin{table*}
\begin{minipage}{13cm}
\caption{Orbital parameters of the planetary system in HU Aqr.}
\begin{tabular}{ll}\hline
Parameters & Values\\\hline\hline
Revised epoch, $\Delta{T_{0}}$(days)& $-1.40(\pm0.56)\times{10^{-4}}$\\
Revised period, $\Delta{P_{0}}$(days)& $+1.59(\pm0.41)\times{10^{-8}}$\\
Rate of the linear decrease, $\beta$(day/cycle) & -4.9($\pm$1.0)$\times10^{-13}$\\
Longitude of the periastron passage (HU Aqr (AB)c), $\omega$(deg) &3.4($\pm0.5$)\\
Periastron passage (HU Aqr (AB)c), T(BJD) &2450299.4($\pm104.1$) \\
Orbital phase (HU Aqr (AB)c), $\varphi$(deg) &
13.8($\pm0.6$)\\\hline
\end{tabular}
\begin{tabular}{lll}
Parameters & HU Aqr (AB)b & HU Aqr (AB)c\\\hline\hline
Light travel-time effect amplitude, $K_A$ and $K_B$ (days) &0.000107(17) & 0.000122(14)\\
Eccentricity, $e_A$ and $e_B$ & 0.0 & $0.51(\pm0.15)$ \\
Orbital period, $P_{A}$ and $P_{B}$ (years) & $6.54(\pm0.01)$ & $11.96(\pm1.41)$\\
$d_A$ and $d_B$ ($i_A=i_B=90^{\circ}$)(AU)& $3.6(\pm0.8)$ & $5.4(\pm0.9)$ \\
Mass function, $f(m_A)$ and $f(m_B)$($M_{\odot}$)& $1.49(\pm0.32)\times{10^{-7}}$ & $0.66(\pm0.13)\times{10^{-7}}$ \\
Projected masses, $M_A\sin{i_A}$ and $M_B\sin{i_B}$($M_{Jup}$) &
$5.9(\pm0.6)$ & $4.5(\pm0.5$)
\\\hline
\end{tabular}
\end{minipage}
\end{table*}

\section{The multiple planetary system orbiting HU Aqr}

\begin{figure}
\begin{center}
\includegraphics[angle=0,scale=.8]{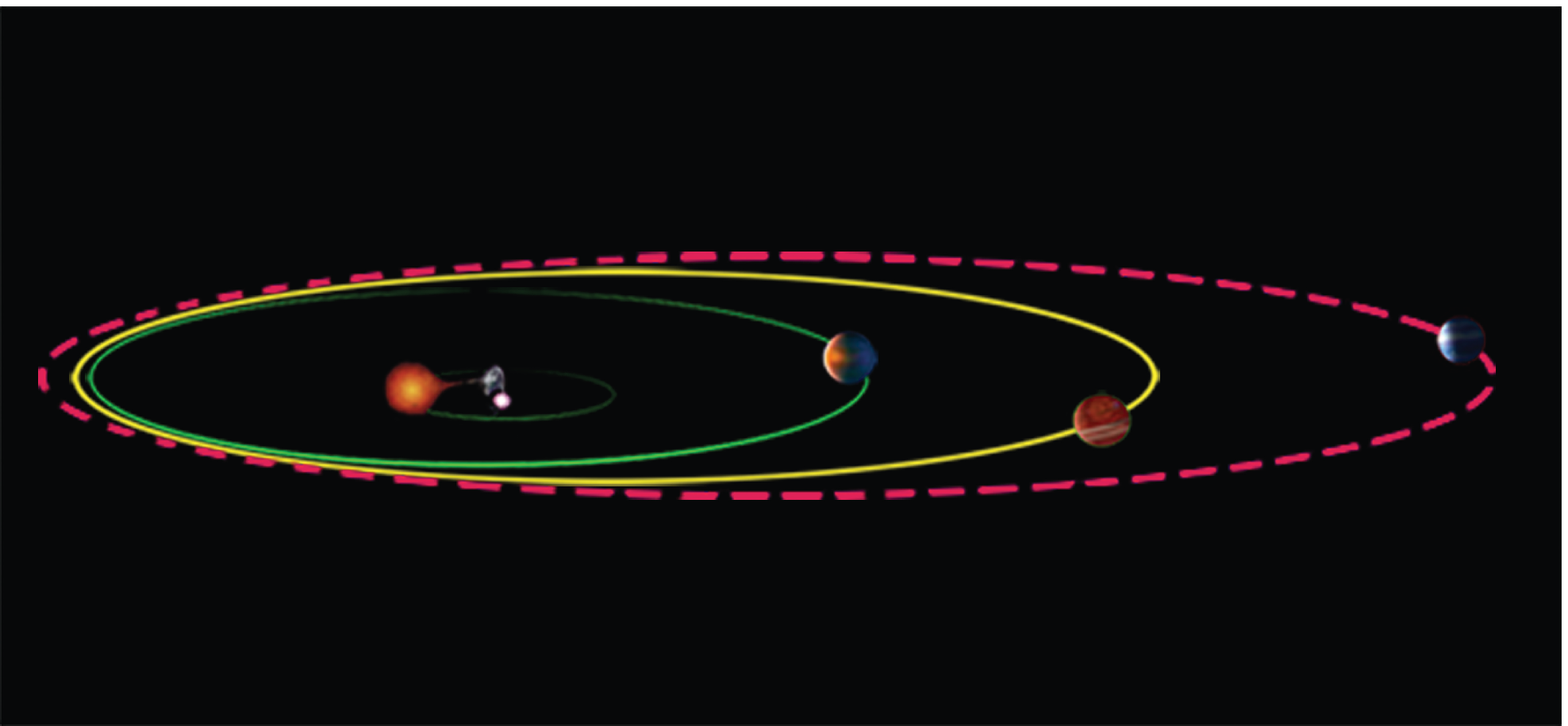} \caption{Sketch of
the eclipsing polar HU Aqr and its cucirmbinary planetary system.
The inner planet (HU Aqr (AB)b) is orbiting in a circular orbit,
while the outer planet (HU Aqr (AB)c) is in an elliptical orbit with
eccentricity of 0.51. The distances of the two planets from the
central binary are 3.6 AU and 5.4 AU, respectively. The red dashed
line refers to the orbit of the predicted third circumbinary planet
(HU Aqr (AB)d).} \end{center}
\end{figure}

To interpret cyclic period variations in close binaries containing
at least one late-type star, a physical mechanism based on
solar-type activity cycles was proposed (Applegate 1992). { In this
mechanism}, a certain amount of angular momentum is periodically
exchanged between the inner and the outer parts of the convection
zone, and { therefore the rotational oblateness of the partly
convective solar-like star, and thus the orbital period, will vary
while} it goes through its active cycles. The secondary star (with a
mass of 0.2\,$M_{\odot}$, where $M_{\odot}$ is the solar mass) in HU
Aqr is a fully convective cool star. It has no differential rotation
and it rotates mainly as a rigid body. Magnetic field in this type
of stars is mainly the axisymmetric large-scale stable field (Donati
et al. 2006, 2008). Applegate's mechanism in such kind of cool stars
{  is generally too weak to explain} the observed amplitudes (e.g.,
Brinkworth et al. 2006). The most plausible explanation of the two
periodic changes is wobbles in the system's barycentre due to the
presences of low-mass companion objects (e.g., Qian et al. 2008;
Beuermann et al. 2011a). As the eclipsing polar moves around the
barycentre of the system, { it is periodically closer to or more
distant from the sun} and the time of mid-egress is cyclically
advanced or delayed.

Adopting the parameters: $M_{WD}=0.88\,M_{\odot}$ and {
$M_2=0.2\,M_{\odot}$} for the white dwarf primary and the red-dwarf
secondary (Schwarz et al. 2009), we obtain: $d_A=3.6$\,AU and
$M_A\sin{i_A}=5.9$\,$M_{Jup}$ for HU Aqr (AB)b and $d_B=5.4$\,AU and
$M_B\sin{i_B}=4.5$\,$M_{Jup}$ for HU Aqr (AB)c, where dA and dB are
the companion-binary separations and 1\,AU is the mean distance
between the Earth and the Sun. When the orbital inclinations ($i_A$
and $i_B$) are low, { the companion objects may be a brown dwarf
(for $4.6^{\circ} \la i_A \la 23.6^{\circ}$ and $3.5^{\circ} \la i_B
\la 17.8^{\circ}$) or even a low-mass star (for $i_A \la
4.6^{\circ}$ and $i_B \la 3.5^{\circ}$), but these situations have a
very low possibility}. Moreover, circumbinary planets were expected
theoretically, at least initially, to have a nearly coplanar orbit
with the central binary (e.g., Bonnell \& Bate 1994). Therefore, by
assuming that the companions are coplanar to the eclipsing polar
($i_A=i_B=85.0^{\circ}$) (e.g., Bridge et al. 2002), they should be
giant planets. From our best fit, HU Aqr (AB)b is moving around the
polar in circular orbit, while HU Aqr (AB)c is in an elliptical
orbit with eccentricity of 0.51. { A sketch of the eclipsing polar
HU Aqr and its circumbinary planetary system is displayed in Fig.
3.}

\section{Discussions and conclusions}

It is well known that there is a Titus-Bode law for solar planets
deduced from the known distances of the main planets (e.g., Poveda
\& Lara 2008). { Although the law predicts that there should be a
planet between Mars and Jupiter, this planet does not exist.} {  It
is interesting to point out that the planet-binary distances of the
two planets of HU Aqr follow} the Titus-Bode law of solar planets
with n=5 and 6. As shown in Fig. 3, HU Aqr (AB)b is at the position
of the belt of asteroids, while HU Aqr (AB)c corresponds to the
position of Jupiter. It is unclear whether these agreements are
fortuitous or there are physical reasons. {  Although
substellar-object systems were reported to be orbiting two post-red
giant branch binary stars, i.e., HW Vir (Lee et al. 2009) and NN Ser
(Beuermann et al. 2011b), the circumbinary planetary system in HU
Aqr is the first one orbiting a polar.}

\begin{figure}
\begin{center}
\includegraphics[angle=0,scale=1.1]{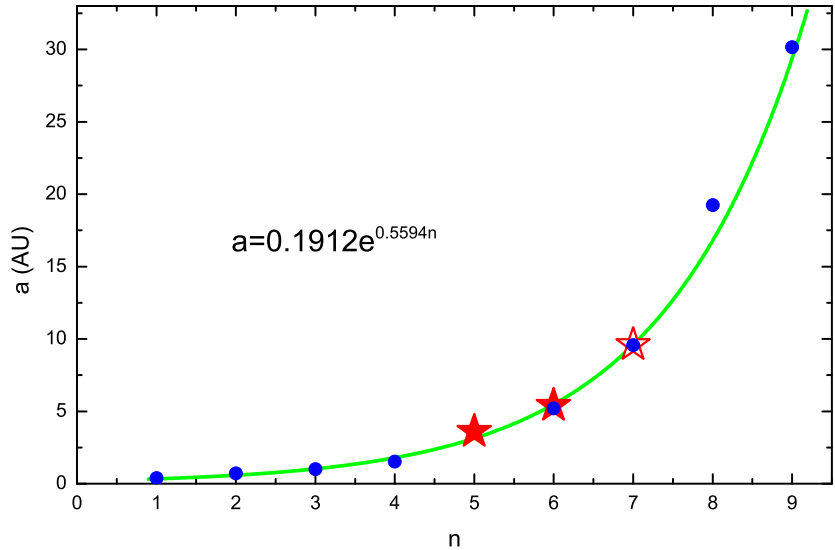}\caption{Titus-Bode
relation for the solar system planets (the green line). The blue
dots refer to the positions of the solar planets, while the red
solid stars are two detected circumbinary giant planets in HU Aqr.
It is found that the two planets follow the Titus-Bode law of solar
planets with n=5 and 6, corresponding the positions of the belt of
asteroids and Jupiter. If the predicted third extrasolar planet in
HU Aqr system follows the same law (n=7, the position of
Soil-Planet), its orbital distance from the central binary should be
9.6 AU (the red open star), another 10 years of observations will
form the full light travel-time effect orbit of the third
circumbinary planet.}
\end{center}
\end{figure}

{ The quadratic term in the O-C diagram (dashed magenta line in the
top panel of Fig. 2)} implies a period decrease at a rate of
$\dot{P}=-0.56\times{10^{-11}}$ (or 1s in about 5,700 years), { that
is half the value obtained by Schwarz et al. (2009). If the observed
period decrease reflects a true angular momentum loss, it would be a
factor 15 larger than the value expected for gravitational
radiation.} This decrease can be explained by magnetic braking
(MB)(Schwarz et al. 2009), { but it is widely accepted that MB is
stopped for fully convictive stars (Rappaport et al. 1983; Spruit \&
Ritter 1983)}. Moreover, the strong magnetic field of the white
dwarf primary may suppress magnetic stellar wind braking in most
polars (King 1994; Wickramasinghe \& Wu 1994). { Therefore, the most
plausible reason} for the observed long-term period decrease is that
it is only a part of a long-period cyclic variation, revealing the
presence of a third planet (HU Aqr (AB)d) in the planetary system. {
We estimate that the planet-binary distance is about 9.6 AU with an
orbital period of 27.1 years. A complete light travel-time effect
orbit of the third circumbinary planet could be obtained in about 10
more years of monitoring.}

It is generally proposed that nonmagnetic CVs (normal CVs with an
accreting disc) and polars like HU Aqr were formed through a
common-envelope (CE) evolution where the secondary star orbits
inside the red giant photosphere of the white dwarf progenitor (King
1994; Kolb 1995). The circumbinary planetary system in HU Aqr may
have been formed in a protoplanetary disk and then survived over
long timescales including the CE evolution phase. It is possible
that the HU Aqr system may have been born with a considerable number
of planets orbiting a star like the Sun. The inner planets spiraled
in the CE after the original central star evolves into a red giant.
During the CE phase, the less massive the spiraled object is, the
longer timescale of the CE evolution will be. Therefore, if one of
the spiraled planets survived, it will have enough time to accrete
so much material to form a low-mass companion. At the same time, the
masses of the other planets were increased due to accreting.
Finally, a polar system with a few relic giant planets was formed.
Similar physical mechanism has been proposed to explain the
formation of the low-mass companions in the two very hot subdwarf
stars HW Vir and AA Dor (Heber 2009; Rauch 2000).

To date, over 30 multiple exoplanet systems were discovered after
the first detection 18\,years ago (Wolszczan \& Frail, 1992). Most
of them are orbiting around single solar-type stars (E.g., Wright
2010). It is possible that some will evolve into polar-planetary
systems. The observational fact that the inner planet (HU Aqr (AB)b)
is more massive than the outer one (HU Aqr (AB)c) supports this
formation process. Moreover, the two planets following the
Titus-Bode law of solar planets appears to be further evidence.
Therefore, most probably, HU Aqr is an offspring of a single-stellar
multiple planetary system. { However, it should be pointed out that
a WD mass of 0.88\,$M_{\odot}$ in HU Aqr suggests a progenitor much
more massive than the Sun. From the initial-to-final mass relation,
the mass of the progenitor should be about 5$\,M_{\odot}$ (Weidemann
2000; Catal\'{a}n et al. 2008a, b).} { The two planets in HU Aqr and
the one in the other polar DP Leo are} more massive than the solar
planets (Qian et al. 2010; Beuermann et al. 2011a). They may have
accreted a large amount of material during the evolution of CE. {
The question is that it is unclear on which conditions a planet will
survive and gain mass. Another possibility is that the circumbinary
planets in HU Aqr are second generation planets that originated from
a disk formed in the ejected envelope, as discussed by Perets
(2010).}

\section*{Acknowledgments}
This work is partly supported by Chinese Natural Science Foundation
(No.10973037, No.10903026, and No.11003040), the National Key
Fundamental Research Project through grant 2007CB815406, the Yunnan
Natural Science Foundation (No. 2008CD157), and by West Light
Foundation of the Chinese Academy of Sciences. New CCD photometric
observations of the system were obtained with the 2.4-m telescope in
the Lijiang station of Yunnan Astronomical Observatory. The authors
thank the referee for those useful comments and suggestions that
help to improve the original manuscript.

\end{document}